\documentclass[longauth,usenames,dvipsnames]{aa} 

\usepackage{natbib}
\bibpunct{(}{)}{;}{a}{}{,} 
\usepackage{graphicx}

\usepackage{siunitx}
\usepackage{txfonts}
\usepackage{threeparttable}

\usepackage{hyperref}
\usepackage{tabularx}
\usepackage{dcolumn}  
\usepackage{collcell}
\usepackage{xcolor}
\hypersetup{colorlinks=true, citecolor=BlueGreen, linkcolor=BlueGreen, urlcolor=BlueGreen}

\newcolumntype{Y}{>{\centering\arraybackslash}X}
\newcolumntype{Z}{>{\raggedleft\arraybackslash}X}
\newcolumntype{d}[1]{D{.}{.}{#1}}    
\usepackage{booktabs}
\usepackage{ulem}

\begin{document}

   \title{The large Trans-Neptunian Object 2002~TC$_{302}$   from combined stellar occultation, photometry and astrometry data}

   \titlerunning{2002~TC$_{302}$ occultation}

   \author{J. L. Ortiz\inst{1}
            \and P. Santos-Sanz\inst{1}
            \and B. Sicardy\inst{2}
            \and G. Benedetti-Rossi\inst{2,3}
            \and R. Duffard\inst{1}
            \and N. Morales\inst{1}
            \and F. Braga-Ribas\inst{4,5,3}
            \and E. Fern\'andez-Valenzuela\inst{6}
            \and V. Nascimbeni\inst{7,8} 
            \and D. Nardiello\inst{7,8,9}
            \and A. Carbognani\inst{10}
            \and L. Buzzi\inst{11} 
            \and A. Aletti\inst{11}%
            \and P. Bacci\inst{12}%
            \and M. Maestripieri\inst{12}%
            \and L. Mazzei\inst{12}%
            \and H. Mikuz\inst{13,14}%
            \and J. Skvarc\inst{13}%
            \and F. Ciabattari\inst{15}%
            \and F. Lavalade\inst{16}%
            \and G. Scarfi\inst{17}%
            \and J. M. Mari\inst{18}%
            \and M. Conjat\inst{19}%
            \and S. Sposetti\inst{20}%
            \and M. Bachini\inst{21}%
            \and G. Succi\inst{21}%
            \and F. Mancini\inst{21}%
            \and M. Alighieri\inst{21}%
            \and E. Dal Canto\inst{21}%
            \and M. Masucci\inst{21}%
            \and M. Vara-Lubiano\inst{1}%
            \and P. J. Guti\'errez\inst{1}%
            \and J. Desmars\inst{22,23}%
            \and J. Lecacheux\inst{2}%
            \and R. Vieira-Martins\inst{5,3}%
            \and J. I. B. Camargo\inst{5,3}%
            \and M. Assafin\inst{24,3}%
            \and F. Colas\inst{23}%
            \and W. Beisker\inst{25}%
            \and R. Behrend\inst{26}%
            \and T. G. Mueller\inst{27}%
            \and E. Meza\inst{2}%
            \and A. R. Gomes-Junior\inst{28,3}
            \and F. Roques\inst{2}%
            \and F. Vachier\inst{23}%
            \and S. Mottola\inst{29}%
            \and S. Hellmich\inst{29}%
            \and A. Campo Bagatin\inst{30,31}
            \and A. Alvarez-Candal\inst{5,31}%
            \and S. Cikota\inst{32}%
            \and A. Cikota\inst{33}%
            \and J. M. Christille\inst{34}%
            \and A. P\'al\inst{35}%
            \and C. Kiss\inst{35,36}%
            \and T. Pribulla\inst{37,38,39}%
            \and R. Kom\v{z}\'{\i}k\inst{37}%
            \and J. M. Madiedo\inst{1}
            \and V. Charmandaris\inst{40,41}
            \and J. Alikakos\inst{40}%
            \and R. Szakáts\inst{35}%
            \and A. Farkas-Tak\'acs\inst{35,42}%
            \and E. Varga-Vereb\'elyi\inst{35}%
            \and G. Marton\inst{35}%
            \and A. Marciniak\inst{43}%
            \and P. Bartczak\inst{43}%
            \and M. Butkiewicz-B\c{a}k\inst{43}%
            \and G. Dudzi\'nski\inst{43}%
            \and V. Al\'{i}-Lagoa\inst{27}
            \and K. Gazeas\inst{43}%
            \and N. Paschalis\inst{44}%
            \and V. Tsamis\inst{45}%
            \and J. C. Guirado\inst{46}%
            \and V. Peris\inst{46}%
            \and R. Iglesias-Marzoa\inst{47,48}%
            \and C. Schnabel \inst{25,49}%
            \and F. Manzano \inst{49}%
            \and A. Navarro \inst{49}%
            \and C. Perell\'o \inst{25,49}%
            \and A. Vecchione \inst{50}%
            \and A. Noschese \inst{50}%
            \and L. Morrone \inst{51}%
          }


        \institute{Instituto de Astrof\'{i}sica de Andaluc\'{i}a, IAA-CSIC, Glorieta de la Astronom\'{i}a s/n, 18008 Granada, Spain,
        \and LESIA, Observatoire de Paris, Universit\'e PSL, CNRS, Sorbonne Universit\'e, Universit\'e de Paris, 5 place Jules Janssen, 92195 Meudon, France
         \and  Laboratório Interinstitucional de e-Astronomia - LIneA, Rua Gal. José Cristino 77, Rio de Janeiro, RJ, 20921-400, Brazil, 
         \and  Federal University of Technology-Paran\'a (UTFPR / DAFIS), Curitiba, Brazil, 
        \and  Observat\'orio Nacional/MCTIC, Rio de Janeiro, Brazil, 
        \and  Florida Space Institute, University of Central Florida, 12354 Research Parkway, Partnership 1, Orlando, FL, USA, 
        \and  Dipartimento di Fisica e Astronomia, 'G. Galilei', Universita degli Studi di Padova, Padova, Italy, 
        \and  INAF-Osservatorio Astronomico di Padova, Padova, Italy, 
        \and Aix Marseille Univ, CNRS, CNES, LAM, Marseille, France
        \and INAF - Osservatorio di Astrofisica e Scienza dello Spazio, Bologna, Italy,
        \and  Schiaparelli Astronomical Observatory, Varese, Italy, 
        \and  Astronomical Observatory San Marcello Pistoiese CARA Project, Italy, 
        \and  Crni Vrh Observatory, Crni Vrh nad Idrijo, Slovenia, 
        \and  Faculty of Mathematics and Physics, University of Ljubljana, Slovenia, 
        \and  Osservatorio Astronomico di Monte Agliale, Lucca, Italy, 
        \and  83560 Vinon sur Verdon, France, 
        \and  Observatorio Astronomico Iota-Scorpii, La Spezia, Italy, 
        \and  06410 Biot, France, 
        \and  Observatoire de la C\^ote d'Azur, France, 
        \and  6525 Gnosca, Switzerland, 
        \and  Osservatorio Astronomico di Tavolaia, Pisa, Italy, 
        \and Institut Polytechnique des Sciences Avanc\'ees IPSA, 63 boulevard de Brandebourg, F-94200 Ivry-sur-Seine, France,
        \and Institut de M\'ecanique C\'eleste et de Calcul des \'Eph\'em\'erides, IMCCE, Observatoire de Paris, PSL Research University, CNRS, Sorbonne Universit\'es, UPMC Univ Paris 06, Univ. Lille, 77 Av. Denfert-Rochereau, F-75014 Paris, France,
        \and Observat\'orio do Valongo/UFRJ, Rio de Janeiro, Brazil, 
        \and International Occultation Timing Association - European Section (IOTA-ES), Germany, 
        \and Observatoire de Geneve, Sauverny, Switzerland,
        \and  Max Planck Institut f\"ur extraterrestrische Physik (MPE), Garching, Germany, 
        \and UNESP - São Paulo State University, Grupo de Dinâmica Orbital e Planetologia, Guaratinguetá, SP, 12516-410, Brazil 
        \and  German Aerospace Center (DLR), Institute of Planetary Research, Berlin, Germany, 
        \and DFISTS, Universidad de Alicante,
        \and IUFACyT, Unversidad de Alicante,
        \and  University of Zagreb, Faculty of Electrical Engineering and Computing, Unska 3, 10000 Zagreb, Croatia, 
        \and  Physics Division, E.O. Lawrence Berkeley National Laboratory, 1 Cyclotron Road, Berkeley, CA, 94720 USA, 
        \and  Astronomical Observatory of the Autonomous Region of the Aosta Valley, 
        \and Konkoly Observatory, Research Centre for Astronomy and Earth Sciences, Konkoly-Thege Mikl\'os \'ut 15-17, H-1121 Budapest, Hungary,
        \and ELTE E\"otv\"os Lor\'and University, Institute of Physics, P\'azmány P\'eter s\'et\'any 1/A, H-1117 Budapest, Hungary,
        \and  Astronomical Institute, Slovak Academy of Sciences, Tatransk\'a Lomnica, Slovakia, 
        \and MTA-ELTE Exoplanet Research Group, 9700 Szombathely, Szent Imre h. u. 112, Hungary,
        \and ELTE Gothard Astrophysical Observatory, 9700 Szombathely, Szent Imre h. u. 112, Hungary,
        \and  Institute for Astronomy, Astrophysics, Space Applications \& Remote Sensing, National Observatory of Athens, Athens, Greece, 
        \and  University of Crete, Department of Physics, Heraklion, Greece, 
        \and E\"otv\"os Lor\'and University, Faculty of Science, P\'azmány P\'eter s\'et\'any 1/A, H-1117 Budapest, Hungary,
        \and  Astronomical Observatory Institute, Faculty of Physics, A. Mickiewicz University, Poznan, Poland, 
        \and  Nunki Observatory, 
        \and  Ellinogermaniki Agogi Observatory, Greece, 
        \and  Observatorio Astron\'{o}mico, Universidad de Valencia, Valencia, Spain, 
        \and  Centro de Estudios de F\'{i}sica del Cosmos de Arag\'{o}n, Teruel, Spain, 
        \and  Dpto. de Astrof\'{i}sica, Universidad de La Laguna, Tenerife, Spain, 
        \and  Agrupaci\'o Astron\'omica de Sabadell, Barcelona, Spain, 
        \and Astrocampania, Osservatorio Salvatore di Giacomo, Agerola (NA), Italy, 
        \and via Radicosa 44, 80051 Agerola, Italy,
        \and The 2002~TC$_{302}$ collaboration
        }

   \date{{Received  XX, 2020; accepted XXX, 2020}}

 
  \abstract
   {Deriving physical properties of Trans-Neptunian Objects is important for the understanding of our solar system. This requires observational efforts and the development of techniques suitable for these studies.}
   {Our aim was to characterize the large Trans-Neptunian Object 2002~TC$_{302}$ }
   {Stellar occultations offer unique opportunities to determining key physical properties of Trans-Neptunian Objects. On 28$^{th}$ January 2018, the large Trans-Neptunian Object 2002~TC$_{302}$ occulted a m$_v \sim $ 15.3 star with designation 593-005847 in the UCAC4 stellar catalog, corresponding to Gaia source 130957813463146112. Twelve positive occultation chords were obtained from Italy, France, Slovenia and Switzerland. Also, four negative detections were obtained near the north and south limbs. This represents the best observed stellar occultation by a Trans-Neptunian Object other than Pluto, in terms of the number of chords published thus far. From the twelve chords, an accurate elliptical fit to the instantaneous projection of the body, compatible with the near misses, can be obtained.}
   {The resulting ellipse has major and minor axes of 543 $\pm$ 18 \si{\kilo\meter} and 460 $\pm$ 11 \si{\kilo\meter}, respectively, with a position angle of 3 $\pm$ 1 degrees for the minor axis. This information, combined with rotational light curves obtained with the 1.5-m telescope at Sierra Nevada Observatory and the 1.23-m telescope at Calar Alto observatory,  allows us to derive possible three-dimensional shapes and density estimations for the body, based on hydrostatic equilibrium assumptions. The effective diameter in equivalent area is around 84 \si{\kilo\meter} smaller than the radiometrically derived diameter using thermal data from Herschel and Spitzer Space Telescopes. This might indicate the existence of an unresolved satellite of up to $\sim$ 300 \si{\kilo\meter} in  diameter, to account for all the thermal flux, although the occultation and thermal diameters are compatible within their error bars given the considerable uncertainty of the thermal results. The existence of a potential satellite also appears to be consistent with other ground-based data presented here. From the effective occultation diameter combined with absolute magnitude measurements we derive a geometric albedo of 0.147 $\pm$ 0.005, which would be somewhat smaller if 2002~TC$_{302}$ has a satellite. The best occultation light curves do not show any signs of ring features or any signatures of a global atmosphere.}
   {}


    \maketitle
%
\section{Introduction}
\label{Introduction}

 Trans-Neptunian Objects (TNOs) are thought to be among the least evolved relics of the solar system formation, residing in the outer parts of the solar system, where the influence of the Sun is less severe than in the inner parts of the solar system. Thus, these icy objects are very important bodies that carry plenty of information on the physical and dynamical processes that shaped our solar system. Therefore, they are key bodies for our understanding of the formation and evolution of the solar system.
According to \cite{Fernandez2020} ``we are at the beginning of the exploration of the Trans-Neptunian region so we look forward to new and
important discoveries that will very likely revolutionize our current view of how the solar system formed and evolved''.

At the time of this writing (February 2020) there are 2416 TNOs (including Pluto), 1085 Scattered Disc Objects (SDOs) plus Centaurs, and 24 Neptune Trojans as listed by the Minor Planet Center\footnote{\url{https://www.minorplanetcenter.net/iau/lists/MPLists.html}.}. 
In order to study these objects, there are many different observational strategies. Among the tools to study TNOs, stellar occultations offer the most powerful means of observing them from the ground  to determine key physical properties such as size and shape. Through stellar occultations, sizes and shapes  with kilometric accuracy can be derived, as well as accurate geometric albedos (when the data are combined with reflected light measurements). This technique is specially fruitful in combination with thermal measurements and modeling \citep{Muller2018}. Finding potential atmospheres on them is also theoretically possible through occultations. Besides, after the recent discovery of a dense ring around Haumea \citep{Ortiz2017} in the context of the previous findings of a ring system around the centaur Chariklo \citep{Braga-Ribas2014}, and a structure in Chiron closely resembling that of Chariklo \citep{Ortiz2015, Sickafoose2019}, there is even more interest in the field of stellar occultations by TNOs, as more rings can potentially be found in the Trans-Neptunian region. 


However, the process from predicting to observing stellar occultations by TNOs is complex and has lots of difficulties. As a result, most of the positive occultation detections thus far have been made from single sites, which gives only limited information \cite[for a review see, e.g.,][]{Ortiz2020}. 
Fortunately, there have been several cases with multichord detections, from which plenty of information was retrieved and published \cite[e.g.,][]{Elliot2010,Sicardy2011,Ortiz2012,Braga-Ribas2013,Benedetti2016,Schindler2017,DiasOliveira2017,Ortiz2017,Leiva2017,Benedetti2019}. 
Here we present the observations and the results of the stellar occultation by 2002~TC$_{302}$ on 28$^{th}$ January 2018, which set a record in the number of detections, together with additional information on rotational light curves and time series astrometry to try to put together a coherent picture of this body. The results are interpreted in the context of other bodies of similar size and features.


\section{Occultation predictions}
\label{Occultation predictions}

The occultation by 2002~TC$_{302}$ on 28$^{th}$ January 2018 was predicted within our program of physical characterization of TNOs by means of stellar occultations. Important international efforts in this regard are currently being coordinated within the framework of the Lucky Star project\footnote{\url{http://lesia.obspm.fr/lucky-star/}} and the observations of this event were organized in that context of collaboration. The prediction was made in different steps with different star catalogs. We used the  HSOY \citep{Altmann2017} and UCAC5 \citep{Zacharias2017} catalogs because the Gaia DR2 catalog \citep{Lindegren2018}  did not exist until April 2018 and Gaia DR1 did not have information on proper motions, whereas UCAC5 and HSOY contained that information for a bright subset of Gaia stars. Additionally, we used different orbit solutions for the TNO (from the AstOrb\footnote{\url{https://asteroid.lowell.edu/main/astorb}}, MPCORB\footnote{\url{https://minorplanetcenter.net/iau/MPCORB.html}}, JPL Horizons\footnote{\url{https://ssd.jpl.nasa.gov/?horizons}}, AstDys\footnote{\url{https://newton.spacedys.com/astdys/}} sites and our own orbit fits). Once the potential occultation seemed favorable enough to be observed, we carried out specific astrometric monitoring runs for 2002~TC$_{302}$ to narrow down the shadow path uncertainty. As it is well known, the positions of stars down to $\sim $20 magnitude in V are now available to a good accuracy, of the order of or below a milliarcsecond (mas) thanks to the Gaia DR2 catalog including proper motion information \citep{Lindegren2018}, but unfortunately, the positions of the TNOs are not known to that level of accuracy. Hence, specific methods have to be developed to solve this problem \citep[see][]{Ortiz2020}. Usually, a careful astrometric monitoring of the target TNO with sufficiently large telescopes and within a few months to a few days prior to the potential occultation combined with a specific analysis of the measurements, is one of the preferred techniques that have yielded positive occultation results. But care must be taken with the existence of satellites, with contamination from faint background stars, and with other effects that may bias the astrometric observations. The main parameters of the star relevant for the occultation are listed in Table \ref{tb:physical_properties}. Note that the angular diameter was estimated according to the expressions given in \citet{vanBelle1999} with colors from the NOMAD catalog \citep{Zacharias2004}.

\begin{figure}
	\includegraphics[width=\columnwidth]{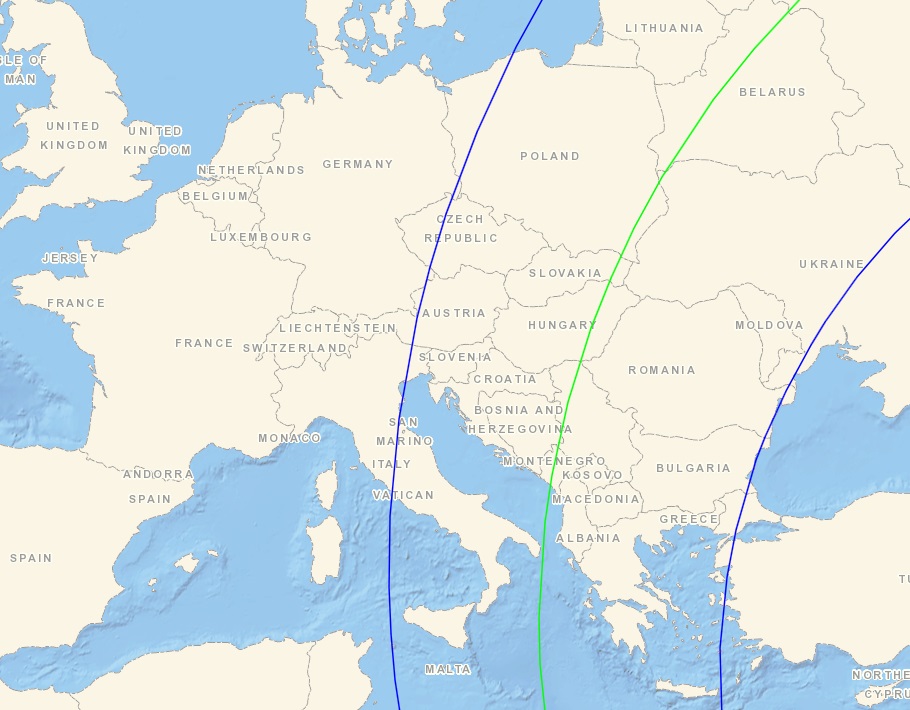}
    \caption{Map showing the shadow path prediction (blue lines) for the stellar occultation by 2002~TC$_{302}$ on 28$^{th}$ January 2018, based on the last set of astrometry measurements obtained prior to the occultation. The green line shows the center of the shadow path. The motion of the shadow is from north to south. The width of the shadow path in the map is 500 \si{\kilo\meter}. The real shadow path of the occultation, obtained after the analysis of the occultation chords, is shown in Figure \ref{observedmap}.}
    \label{predictmap}
\end{figure}

\begin{figure}
	\includegraphics[width=\columnwidth]{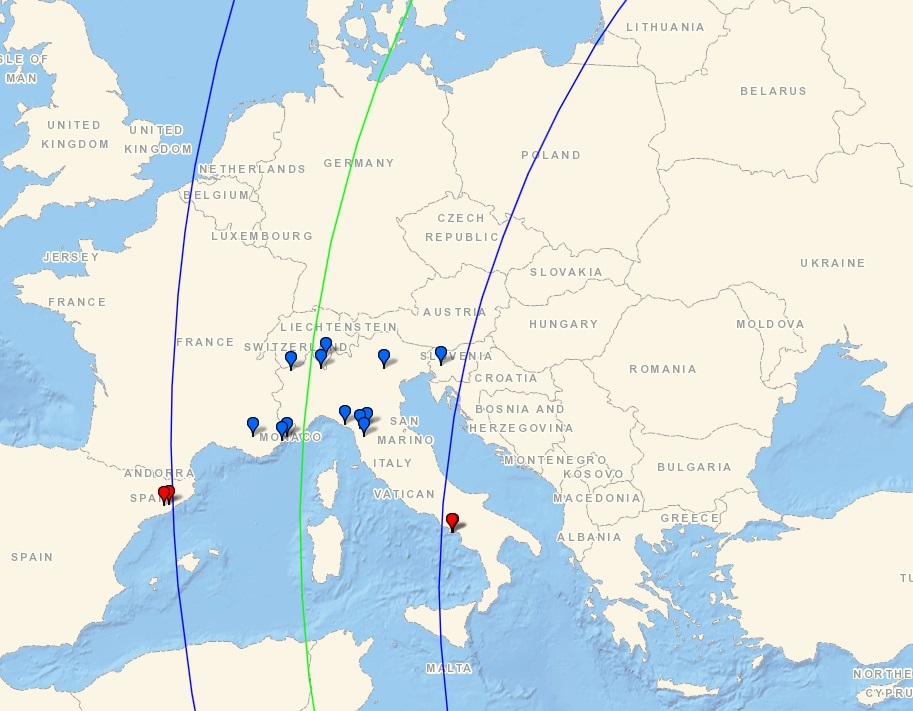}
    \caption{Map showing shadow path (blue lines) reconstructed after the occultation results were obtained. The width of the shadow path used in the map is 500 \si{\kilo\meter}, which is the equivalent-area diameter derived from the occultation. The green line shows the center of the shadow path. The shadow moved from North to South. The blue marks show the observatory sites where the occultation was detected. The two red marks indicate the two observatory sites closest to the shadow, where the event was negative.}
    \label{observedmap}
\end{figure}

From the astrometric monitoring of 2002~TC$_{302}$ carried out in several days of October, November, December 2017 and January 2018 through the use of the Sierra Nevada 1.5m telescope (Granada, Spain) and the Calar Alto 1.2m telescope (Almer\'ia, Spain), we obtained refined predictions with respect to the initial one. From the latest measurements made just a few days prior to the stellar occultation by 2002~TC$_{302}$, the path of the occultation was predicted to be favorable to a large area in Europe (see Figure \ref{predictmap}), although there was some concern that these measurements could be affected by the presence of a potential satellite and the centroid biased to a wrong position. Nevertheless, the prediction made with a specific orbital fit of the NIMA type \citep{Desmars2015} to all the available data indicated a similar path on Earth\footnote{\url{http://lesia.obspm.fr/lucky-star/predictions/single.php?p=3492}}. The final path, reconstructed from the fit to the occultation chords described in subsequent sections,
is depicted in Figure \ref{observedmap}.

The astrometric observations at the 1.5-m telescope of Sierra Nevada Observatory consisted of CCD images taken with the $2{\rm k}\times2{\rm k}$ Versarray camera\footnote{\url{https://www.osn.iaa.csic.es/en/page/ccdt150-camera}}, which has a field of view (FoV) of $7.8\times7.8$ arcmin and a scale of 0.23 arcsec/pixel. The integration time was 400 s and no filters were used in order to maximize the signal to noise ratio of the observations. Typical seeing ranged from 1.3 to 2.5 arcsec. The Calar Alto 1.2m telescope images were acquired using the $4{\rm k}\times4{\rm k}$ DLR camera\footnote{\url{https://www.caha.es/CAHA/Instruments/IA123/index.html}}, which has a FoV of $22\times22$ arcmin and a pixel scale of 0.32 arcsec/pixel. We used the Johnson-Cousins $R$ filter to avoid fringing in the near-IR, which is a known issue in this camera. The typical seeing was around 1.5 arcsec and the used integration times were 400s, with sidereal tracking. At both telescopes, images were obtained using $2\times2$ binning mode and sidereal tracking. All images were bias subtracted and flatfield corrected using bias frames taken each night and using a median of flatfield frames obtained on each observing night (if this was not possible, flatfield frames from a previous night were used).

The images were analyzed with our own software that extracts the sources of the images, solves for the plate constants using a specific astrometric catalog (selected by the user) and then performs the astrometry of the target. The astrometric catalog used was Gaia DR1 because Gaia DR2 had not been released prior to the occultation. Therefore, no proper motion corrections were applied to the reference stars. This process is fully automated, but visual inspection was made to discard images in which the target could be blended with a background star and to discard images with cosmic ray hits near the target. Also, care was taken so that charge bleeding or blooming or ghosts from bright stars in the FoV or any anomalous aspect did not affect the TNO measurements. Online Table 1 lists all the measurements referred to the J2000 equinox. The NIMA prediction also included observations performed at Pico dos Dias Observatory (OPD) in October (18/10/2017) and November (12/11/2017) with the 1.6-m Perkin Elmer Telescope and using the Andor-IKon camera\footnote{\url{http://www.lna.br/opd/instrum/ccd/manual_ikon.pdf}} (pixel scale = 0.180 arcsec/pixel, FoV $=7\times7$ arcsec). The images were calibrated with bias and flatfields taken during the same night and the used exposure times were 180s in Johnson - $I$ filter.

\begin{table*}
	\centering
	\caption{ Main characteristics of the occulted star.}
	\label{tb:physical_properties}
	\begin{tabular}{cccccc}   
		\hline	
		\hline	    
    Gaia ID & RA & Dec & G  & Ang. Diam. & Speed  \\
    &(hh mm ss.s)&(\degr ~\arcmin ~\arcsec)&(mag)&(mas)&(\si{\kilo\meter\per\second})\\\hline
    130957817758443648	& 02 21 49.3853105797	& +28 24 13.439342645  & 15.6 & 0.009 & 4.77  \\\hline
	\end{tabular}
	\tablefoot{Abbreviations are defined as follows: Gaia DR2 identification number (Gaia ID), J2000 coordinates of the star from Gaia DR2 (right ascension and declination, RA and Dec., respectively), G magnitude (G), angular diameter (Ang. Diam.), speed relative to the observer (speed).}
\end{table*}


\section{Occultation observations}
\label{Occultation Observations}

The occultation observations were performed with different telescopes and with various camera setups. The main observational details for the occultation observations at the main sites involved in the campaign are listed in Table \ref{tb:sites}. We list only the sites where the event was positive or where a close miss to the final shadow path was produced (providing at least constraints to the final fit of a shape model). More observatories than those listed in Table \ref{tb:sites} monitored the event, but did not achieve positive results for a variety of reasons (because they were far from the final occultation path or because they were clouded out or had technical problems). Unfortunately, a complete list of all the observatories that participated in the campaign cannot be derived, because there were amateur observers alerted through different internet tools and other procedures that we could not monitor. Nevertheless, we are keeping a registry\footnote{\url{http://asteroidstnos.iaa.es/content/sharedfiles}} of all the observers who report their participation. This registry will be updated whenever new information becomes available.

Most of the observations from the sites in Table \ref{tb:sites} consisted of sequences of images taken with different telescopes and different CCDs or CMOS detectors, as specified in the table. Other observations, indicated also in the table, were acquired in video mode. The video observations required a different analysis as explained in Section \ref{Analysis}. No filters were used to maximize the number of photons received in order to get the highest possible signal to noise ratio. Table \ref{tb:sites} shows the names of the observing sites, their topocentric coordinates (longitude and latitude), the exposure time, the cycle time between consecutive exposures, the diameters of the telescopes, and the detector manufacturers and models.

The observations started typically 15 minutes before the predicted time for the occultation and were finalized around 15 minutes after the event. This was done in order to determine a good base line for the photometric analysis, and to determine its noise level before and after the occultation event. The moon was 90\% full at 52 degrees from the target. This means that considerable sky background affected the observations and therefore the signal to noise ratio of the observations was not as high as it could have been in the absence of moon illumination. Weather conditions were mostly clear in all the observatories except at Tavolaia, where intermittent clouds were present. Nevertheless, all the observing sites achieved enough signal to noise ratio in the images (see last column of Table \ref{tb:times}) so that the occultation brightness drop was clearly detected.

The campaign around this occultation was a major achievement because no stellar occultation by a TNO had ever been observed with so many chords across the main body and with near misses.

\begin{table*}
\sisetup{
table-format = 2.3 ,
table-number-alignment = center ,
}
	\centering
	\caption{Observatories and characteristics of the observations.}
	\label{tb:sites}
	\begin{tabularx}{\textwidth}{p{3cm}p{2cm}c*{3}{S}Y}   
		\hline	
		\hline	    
		Site        &  \multicolumn{1}{Y}{Longitude~(E)}  & Latitude (N)  & \multicolumn{1}{Y}{Exp. time} & \multicolumn{1}{Y}{Cycle time}    & \multicolumn{1}{Y}{Telescope diameter} & Detector \\
		& \multicolumn{1}{Y}{(\degr)} & (\degr) & \multicolumn{1}{Y}{(s)} & \multicolumn{1}{Y}{(s)} & \multicolumn{1}{Y}{(cm)} & \\\hline
		Crni Vrh    & 14.071083     & 45.945833 & 3.4               & 4.995             & 60      & Apogee Alta U9000HC \\
		Asiago      & 11.568806     & 45.849444 & 5.0               & 8.301             & 67     &        Moravian G4-16000LC, KAF-16803    \\
		S. Marcello Pistoiese & 10.803889 & 44.064167 & 3.0         & 3.933             & 60      & Apogee Alta U6  \\
        Tavolaia    & 10.673306     & 43.736500 &           7        &          7.23         & 40      &           ASI 174MM     \\
        Mount Agliale & 10.514944   & 43.995278 & 2.0               & 3.718             & 50      & FLI proline 4710\\
        La Spezia     & 9.853528    & 44.126278 & 4.0               & 7.830             & 40      &       Sbig STXL 6303e          \\
        Gnosca        & 9.024028    & 46.231444 &     2.56              &           2.568        & 28     & video WAT-910HX-RC, ICX429ALL \\
        Varese, Schiaparelli Observatory & 8.770278 & 45.868056 & 2.0 & 5.716           & 84     &  SBIG STX-16803   \\
        Observatoire de Cote d'Azur, Nice & 7.299833 & 43.725806 & 2.2 &           2.2005       & 40      &       ASI 174MM   \\
        Aosta Valley & 7.478333     & 45.789444 &          2.0         &         2.695          & 40      &  FLI1001E \\
        Biot, Nice   & 7.077778     & 43.617222 & 20.0              & 26.463            & 20      & video QSI 583 wsg, KAF 8300 \\
        Vinon sur Verdon & 5.796111 & 43.737778 &          10         &          30.467         & 30      &  Atik 383L, KAF 8300\\\hline\hline
        Near Misses & & & & & \\\hline
        Osservatorio Salvatore Di Giacomo  & 14.564056 & 40.623944 &  2 &            3.4       &    50       &  FLI PL4220, E2V CCD42-40-1-368 \\
        Agerola     & 14.571556     & 40.626083 &          2.5         &           2.5        &     25      & ASI 178M, IMX178 \\
        Sabadell    & 2.090167      & 41.550000 &           2.56        &           2.56        &     50     &  Watec 910HX-RC  \\
        San Esteve  & 1.872528      & 41.493750 &        2           &              2     &  40         &  Point Grey Chameleon3 \\\hline
	\end{tabularx}

\end{table*}


\section{Analysis of the occultation observations}
\label{Analysis}

Synthetic aperture photometry of the occultation star (blended with the TNO) was derived for the sequences of images of the different cameras in order to obtain the light curves for each site. The synthetic aperture photometry results were derived by means of an Interactive Data Language (IDL) code that uses the implementation of the well known \texttt{DAOPHOT} photometry package \citep{Stetson1987}. Also, synthetic aperture photometry was derived for comparison stars close to the target star in the FoV of the cameras, so that sky transparency fluctuations as well as seeing variations could be taken into account. The final light curves were obtained by taking the ratio of the occulted star flux in ADUs to that of a comparison star (or the combination of several comparison stars if this was possible). We carefully monitored the dispersion of the final light curves and chose the synthetic aperture diameters as well as the rest of parameters of the synthetic aperture technique to get the least possible scatter in the photometry. Centroid tracking to recenter the apertures was done for the reference stars but not for the occultation star, whose position was kept fixed with respect to the references. The time of each photometry point was derived from the time stamps in the FITS headers of the images. It must be noted that some cameras inserted header times rounded off to the nearest second or truncated seconds. In these cases, we interpolated times for each point by using linear fits to the times versus frame number, in the same way as described in \citet{Sicardy2011}. It must also be noted that for the Tavolaia observations, there was a problem in the acquisition that prevented from saving the time in the headers of the images, so we could only use the time of the recording of the file to disk as provided by the operating system. Hence, the timing of these observations is more uncertain than the rest, and in fact, we have accounted for this as explained in section \ref{Size and shape}.
The rest of the sites other than Tavolaia used internet NTP time servers to synchronize their image acquisition computers. Even though this technique is capable of providing accuracies of 0.01 s, the way in which different operating systems and camera control software deal with time (and possible shutter opening delays) is often not accurate to the level of 0.1 s. In fact, errors up to a few tenths of a second have been reported in internet-synchronized devices under Windows operating systems \cite[e.g.,][]{BarryGault2015}. As we mention in a paragraph below, these time errors are nevertheless smaller than those arising from the square-well fits to determine ingress and egress times (given the photometric errors and relatively large exposure times as well as large readout times).

The video observations required a specific analysis. We used the Tangra video analysis tool\footnote{\url{http://www.hristopavlov.net/Tangra3/}} to derive the light curves and the timings that came from the video-inserted GPS-based time stamps whose accuracy is in the order of the millisecond once the small time delays of the video cameras (which depend on the manufacturer) are taken into account.
For the instrumental delay Tangra uses the tables provided by
Gerhard Dangl.\footnote{ \url{ http://www.dangl.at/ausruest/vid_tim/vid_tim1.htm}}

\begin{figure*}
\centering
		\includegraphics[width=0.9\hsize]{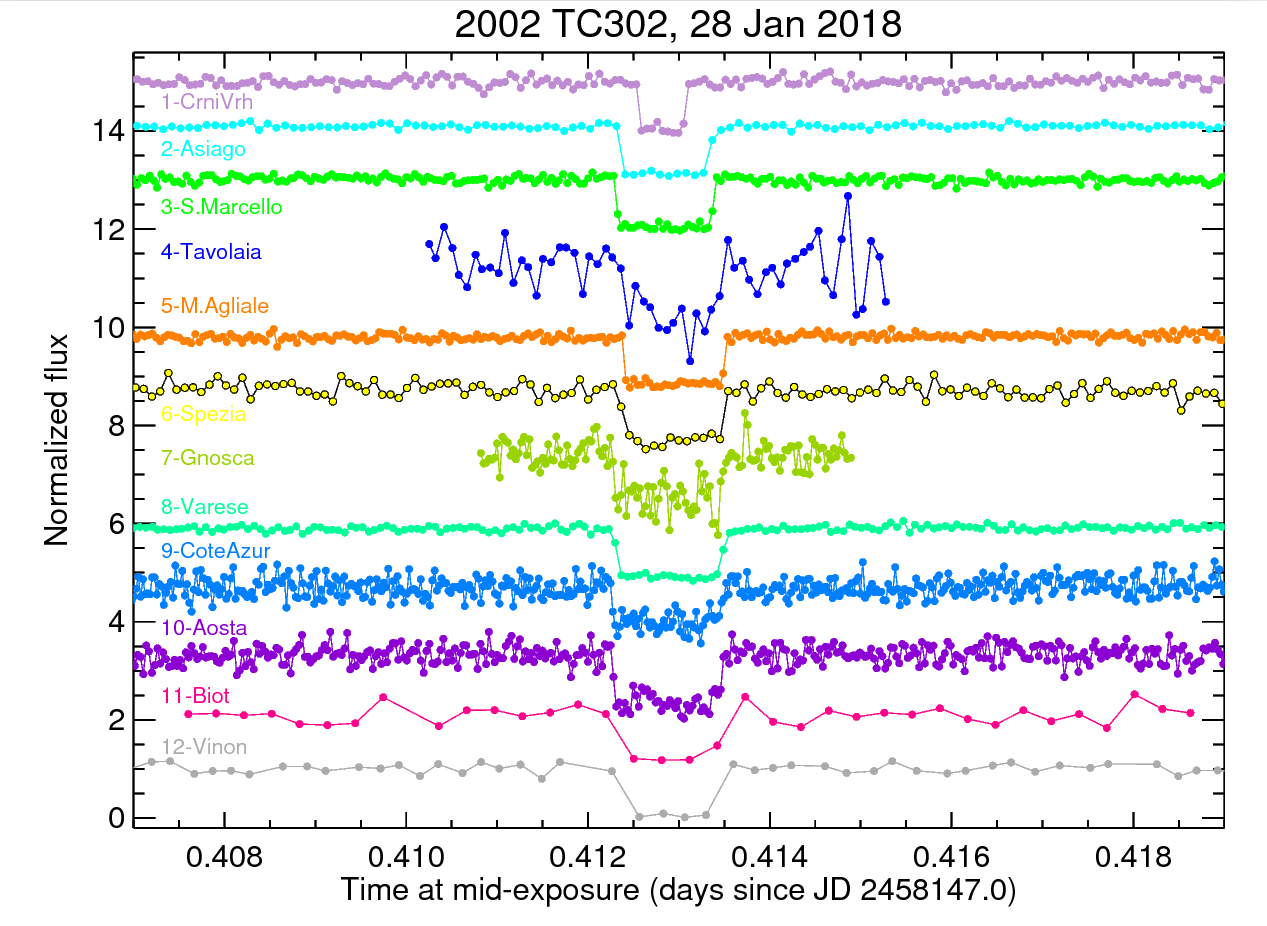}
    \caption{Light curves in normalized flux from all the observatories. The light curves have been displaced in the horizontal axis to account for the different longitudes so that all the occultation drops are visually aligned.}
    \label{observedlightcurves}
\end{figure*}

All the light curves were normalized to the mean brightness level of the blended star and TNO, outside the main occultation event, by dividing the flux by that mean level. Hence, the mean level of the normalized light curve has the value 1 outside the occultation part. The light curves in normalized flux are shown in Figure \ref{observedlightcurves}.
The uncertainties in the fluxes were derived from the photometry using Poisson noise estimations as obtained in the photometry software package \texttt{DAOPHOT}.  Note that all the uncertainties or error bars given throughout the paper are 1$\sigma$. The average values of the theoretical uncertainties were obtained and compared with the standard deviation of the observations. If a departure was present in the theoretical errors compared to the standard deviation, the individual errors were multiplied by the ratio of the observed and the computed standard deviation. This was often needed because the gain values (number of electrons per count on the detector) were not well known for most of the devices. Hence, we assigned a different error bar to each individual measurement (in other words, we did not assume a constant error bar equal to the standard deviation of the measurements). This is particularly relevant for the values at the occultation, when the relative uncertainties are much higher than outside the occultation.

Once all the occultation light curves were derived, we proceeded to fit square well models to the parts of the light curves that showed the occultation. The fits were performed using the same expressions and methodology as in other occultation works \cite[e.g.,][]{Braga-Ribas2014,Benedetti2016,DiasOliveira2017,Ortiz2017,Benedetti2019}. From those fits, the star disappearance and reappearance times were derived and their uncertainties were determined as those values providing fits such that the values of $\chi^2$ were within the interval [$\chi^2_{min}$,$\chi^2_{min} + 1$]. The disappearance and reappearance times and their errors are listed in Table \ref{tb:times}. 


\begin{table*}
\centering
\caption{Disappearance and reappearance times.}
\label{tb:times}
\begin{tabular}{cr@{\ \ $\pm$ \ \ }lr@{\ \ $\pm$ \ \ }lc}
\hline	
\hline
Site    &\multicolumn{2}{c}{Ingress time}   &\multicolumn{2}{c}{Egress time}    & \multicolumn{1}{c}{rms of the} \\
& \multicolumn{2}{c}{(hh:mm:ss.s $\pm$ s.s)} & \multicolumn{2}{c}{(hh:mm:ss.s $\pm$ s.s)} & \multicolumn{1}{c}{normalized flux} \\
\hline
Crni Vrh                            &21:52:51.687 & 1.150       &21:53:35.510 & 0.350       & 0.085 \\
Asiago                              &21:52:43.400 & 1.400       &21:54:08.875 & 0.200       & 0.042 \\
S. Marcello Pistoiese               &21:53:22.837 & 0.475       &21:54:54.051 & 0.163       & 0.062 \\
Tavolaia                            &21:53:36.3 & 2.5           &21:55:08.7 & 2.5           & 0.36 \\
Mount Agliale                       &21:53:23.150 & 0.300       &21:54:57.450 & 0.100       & 0.061 \\
La Spezia                           &21:53:18.850 & 0.500       &21:54:55.550 & 2.400       & 0.139 \\
Gnosca                              &21:52:37.170 & 1.250       &21:54:21.850 & 1.250       & 0.173 \\  
Varese, Schiaparelli Observatory    &21:52:42.585 & 0.100       &21:54:24.775 & 0.100        & 0.049 \\
Observatoire de Cote d'Azur, Nice   &21:53:39.661 & 0.280       &21:55:23.301 & 0.720       & 0.174 \\
Aosta Valley                        &21:52:49.240 & 0.440      &21:54:31.565 & 0.420        & 0.176 \\
Biot, Nice                          &21:53:46.800 & 5.500      &21:55:24.750 & 3.500        & 0.174 \\
Vinon sur Verdon                    &21:53:52.08 & 6.70       &21:55:21.13 & 6.00           & 0.096 \\

\hline
\end{tabular}
\end{table*}

Since the brightness drops at the occultation were all sharp, there is no evidence at all for a global atmosphere in 2002~TC$_{302}$. The square well models provided very satisfactory fits, without any need for incorporating an atmosphere. Specific calculations to derive upper limits for the pressure of putative atmospheres of different compositions would be needed, but such calculations are beyond the scope of this paper because the range of possible atmospheric compositions and temperature profiles is too wide. However, to give an idea, based on similar calculations made for Makemake \citep{Ortiz2012}, Quaoar \citep{Braga-Ribas2013}, and 2003~AZ$_{84}$ \citep{DiasOliveira2017}, for which similar noise levels of the light curves were obtained and given that the sizes of the bodies were similar, we can guess that 2002~TC$_{302}$ lacks a global atmosphere with upper limits of the order of 100 nbar in pressure for N$_2$ and CH$_4$ compositions, as described in the papers mentioned above.

The apparent magnitude of 2002~TC$_{302}$ on 22$^{nd}$ January 2018, measured with respect to the Gaia DR1 G band, turned out to be 20.40 $\pm$ 0.07 mag, using the image set from the 1.5-m Sierra Nevada telescope taken closest to the occultation date. Given that the occulted star has a G magnitude of 15.589 according to the Gaia DR1 catalog, this means that the expected brightness at the bottom of the occultation would be 0.012 in normalized flux.

The derived times of ingress and egress and their uncertainties were the basis to determine the chords of the occultation, once the positions of the TNO were projected in the plane of the sky. The chords were then used for the subsequent step of determining the projected size and shape.


\section{Projected size and shape}
\label{Size and shape}

Since we expect that large TNOs like 2002~TC$_{302}$ should be triaxial ellipsoids or spheroids \citep[e.g.,][]{TancrediFavre2008}, and the two-dimensional projection of these shapes is an ellipse, it makes sense fitting such an ellipse to the extremities of the chords.

\begin{figure*}
\centering
	\includegraphics[width=0.9\hsize]{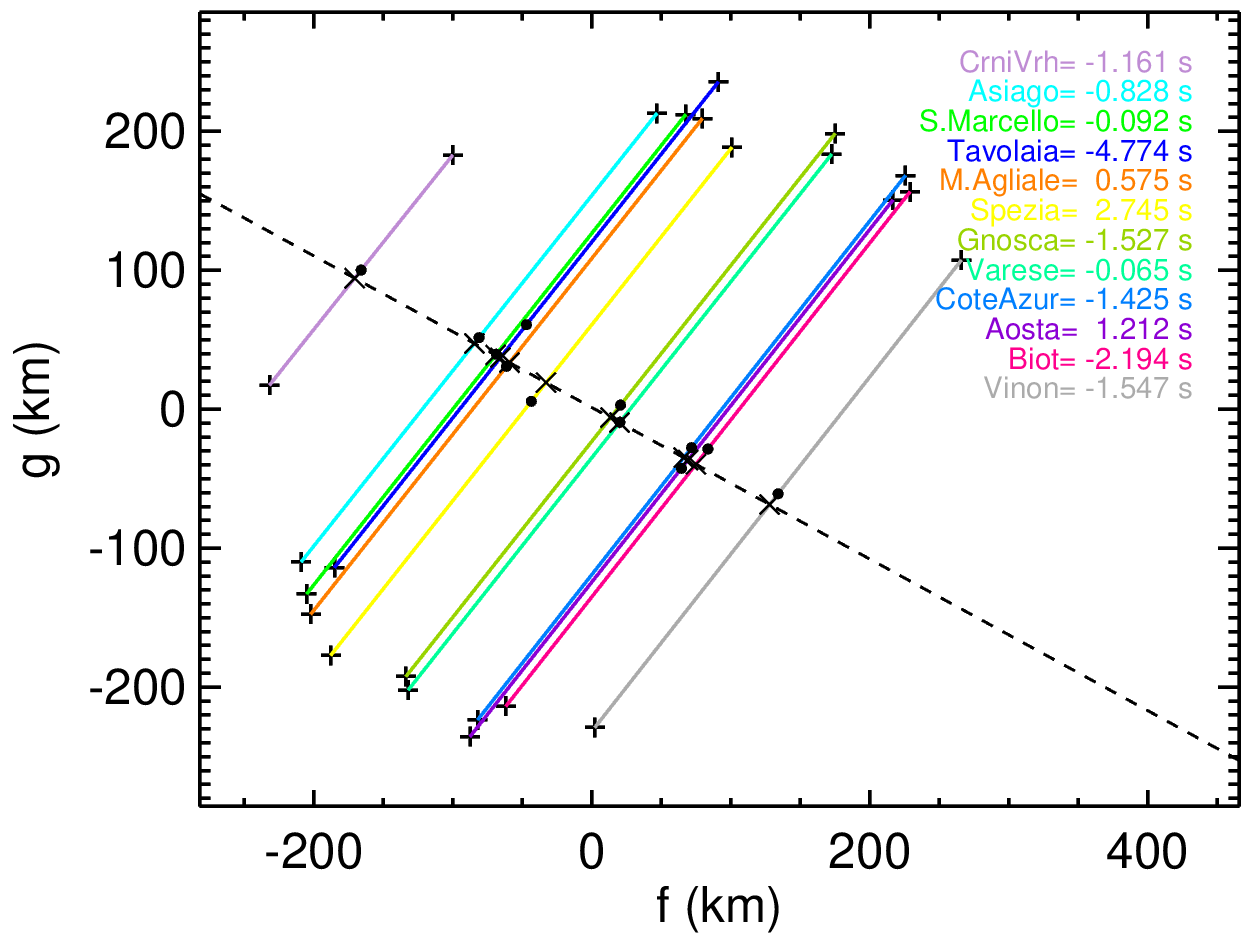}
    \caption{Chords of the occultation in the sky plane and a linear fit to the centers of the chords. North is up, East to the left. Each color represents a different chord as labeled in the insert. The time shifts needed for centers of the chords to be aligned are also shown in the insert. The black dots denote the centers of the chords and the dashed line represents the fit.}
    \label{linearfitchordscenters}
\end{figure*}

Because, as explained in section \ref{Occultation Observations}, the chord from Tavolaia was affected by a considerable uncertainty in time due to the technical issue, we decided to look for the best shifts of the chords that would result in their centers lying on a straight line. We know that this condition must be satisfied by an ellipse and we have the a priori knowledge that the projected shape of 2002~TC$_{302}$ must be an ellipse. Hence, by shifting the chords in this way we make sure that we get the shape that best matches with the theoretical one. Note that, in addition to small topographic relief, which can cause small decentering of the chords, unexplained time shifts of up to 27s have been reported before \citep{Elliot2010}, and \citet{Braga-Ribas2013} also identified smaller but noticeable shifts, so the possibility to shift the chords must be considered. Hence, a linear fit of the chord centers (with each chord center weighted by its nominal uncertainty) was performed (see Figure \ref{linearfitchordscenters}). The shifts were determined from the residuals of the straight line fit.

The ellipse fit was carried out following the same methods as in previous occultations, \cite[e.g.,][]{Braga-Ribas2013,Braga-Ribas2014,Benedetti2016,DiasOliveira2017,Ortiz2017,Benedetti2019}. The fitted parameters are the center of the ellipse, the semiaxes and the tilt angle of the ellipse.

\begin{figure*}
\centering
	\includegraphics[width=0.9\hsize]{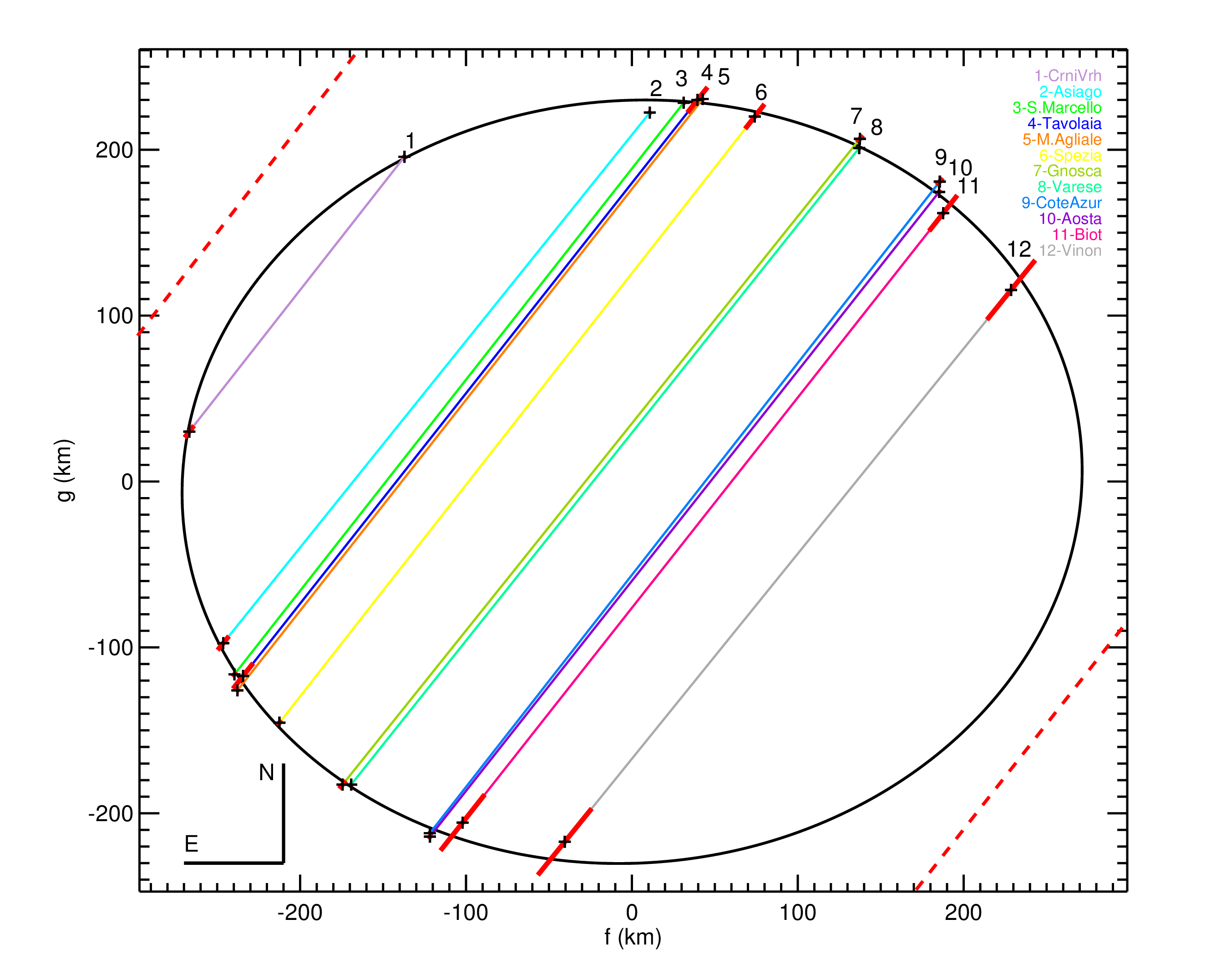}
    \caption{Chords of the occultation in the plane of the sky and elliptical fit to the chords. The color coding is the same as that in Figures \ref{observedlightcurves} and \ref{linearfitchordscenters}. The red segments at the chord extremities show the uncertainties from the ingress and egress fits.  Note the asymmetry between ingress and egress timing uncertainties in tracks 1, 2, and 6. This is due to operational overheads of the detectors and if the ingress (or the egress) takes place during one of these periods, the uncertainty is larger than that solely due to the photometry noise}. The near misses at Sabadell and Agerola are indicated as dashed red lines (easternmost and westernmost respectively).
    \label{fittochords}
\end{figure*}

The resulting ellipse fit is illustrated in Figure \ref{fittochords}. The axes of the ellipse are 543.2 $\pm$ 18 \si{\kilo\meter} and 459.5 $\pm$ 11 \si{\kilo\meter}, with a position angle of $3\pm1$ degrees. In this case the errors were determined by using the same procedure as in previous occultation works \cite[e.g.,][]{Braga-Ribas2013,Braga-Ribas2014,Benedetti2016,DiasOliveira2017,Ortiz2017,Benedetti2019}. The equivalent diameter in projected area is 499.6 \si{\kilo\meter}.

Whether the three dimensional shape of 2002~TC$_{302}$ is a triaxial ellipsoid (with $a>b>c$, and $a$, $b$, $c$ being the semiaxes of the body) or an oblate spheroid (with $a=b>c$) is something that cannot be determined from an occultation alone (several occultations at different rotational phases would be needed, or rotational light curves should be obtained to complement the occultation information). In our case we combined rotational light curves with the occultation. We discuss about the two possible shapes in the next sections.

\section{Light curves of 2002~TC$_{302}$ to determine the rotational period}
\label{Rotational light curves}

The occultation-derived ellipse is just an instantaneous projection of the three-dimensional shape. Therefore, in order to interpret the occultation results, determining the rotation period and rotational light curve is important. If the rotational light curve is double peaked, then it is very likely that the object is a triaxial body, although it could also be an oblate spheroid with a large irregularity. There are also cases of bodies with an oblate shape that present double-peaked rotational light curves arising from albedo variability on their surfaces. A notable example of this is the dwarf planet Ceres, whose oblate shape is well known from stellar occultations \citep[e.g.,][]{GomesJunior2015} and the DAWN spacecraft visit \citep[e.g.,][]{Russel2016}, while it exhibits a low-amplitude double-peaked rotational light curve due to albedo features \cite[e.g.,][]{Chamberlain2007}. 

\cite{Thirouin2012} obtained in 2010 the rotational light curve of 2002~TC$_{302}$ by using the 1.5-m telescope at Sierra Nevada Observatory. That rotational light curve appeared to be single-peaked with a period of 5.41 h, although periods of 4.87 and 6.08 \si{\hour} were also possible.
Unfortunately, the amplitude of the variability was not high (only $0.04\pm0.01$ mag), and in such cases, the confidence in the determined rotation period is difficult to asses. In fact, a clear example of this problem is illustrated with the case of the dwarf planet Makemake, for which 24h-aliases can have very similar spectral power or even higher spectral power than the true rotation period when using data with noise levels similar to the amplitude of the variability.  For the dwarf planet Makemake, potential periods of 11.24, 11.41, and 20.57 \si{\hour} were initially identified in the periodogram derived shortly after its discovery \citep{Ortiz2007}, but later on, it was found that the 11.41 \si{\hour} period was the closest 24-h alias of a preferred single-peaked period of 7.77 \si{\hour} \citep{HeinzeDeLahunta2009}. However, the most recent work on Makemake photometry, using a very large time series, indicates that a double-peaked rotational light curve is favored and the true rotation period is twice the 11.41 \si{\hour} period \citep{Hromakina2019}. Therefore, finding the correct rotation period and rotational light curve of bodies with variability of low amplitude is considerably difficult and there is a clear bias to detect shorter periods, which are much easier to detect than longer periods \citep{Sheppard2008}.

Hence, it was important to analyze more data on 2002~TC$_{302}$ to try to shed light on its rotational light curve and rotation period. Within our program of physical characterization of TNOs, we had observed 2002~TC$_{302}$ in 2014 and 2016 with the 1.5-m telescope at Sierra Nevada Observatory and with the 1.2-m telescope at Calar Alto observatory in specific photometry runs. After the successful occultation we also carried out specific runs in 2018 and 2019. 
The observations were performed in the same way and with the same instruments as described in section \ref{Occultation predictions}.
The methods and tools used to extract the photometry were the same as those explained in \cite{Fernandez-Valenzuela2019}. A total of 875 measurements were obtained, which can be found as on-line material.

The observing runs in 2019 were the most complete ones in terms of the number of consecutive observation nights and the coverage in number of hours per night (and also in terms of signal to noise ratio). In those runs we observed the target for 7 to 9 nights in a row, and most of the nights covered more than 8 hours and up to 10 hours on the target. Using the photometry from those observation nights in 2019 it was already clear that a rotation period of $\sim5.41$ h was not seen in the data. The light curves folded to that period (or values close to it) did not show convincing variability and low dispersion. Given that most of the nights had data in time spans longer than 8 h and no clear variability was seen, it appeared that longer periods than 8 h would be favored. 

There is additional reasoning to favor that the preferred rotation period for 2002~TC$_{302}$ should be longer than $\sim5.4$ \si{\hour}. 
According to \cite{TancrediFavre2008} the typical size for which hydrostatic equilibrium would be expected in icy bodies like the TNOs is well below that of 2002~TC$_{302}$. The equilibrium shapes can be Maclaurin or Jacobi shapes \citep{Chandrasekhar1987}. 
The minimum density that a Maclaurin body with 5.4 \si{\hour} rotation could have is $\sim$ 1150 \si{\kilo\gram\per\meter\cubed}. For this density the axial ratio $a/c$ is larger than 2.5 using the Maclaurin sequence. Hence, in order to give rise to the projected axial ratio $a/c$ of 1.18 seen in the occultation, the aspect angle would have to be extremely low, which is very unlikely, and besides, the density is considerably high for a TNO of this size \citep[see e.g. density vs. size plots in][]{Grundy2015,BiersonandNimmo2019,Grundy2019}. Another possibility is a Jacobi body; however, given the short rotation period and axis ratio detected from the occultation, this would require an even larger density.
For more plausible densities, well below 1150 \si{\kilo\gram\per\meter\cubed}, there is no hydrostatic equilibrium shape possible for a homogeneous body with rotation period of 5.4 \si{\hour}. There is the possibility that 2002~TC$_{302}$ has adopted an oblate spheroid shape but that the true density could be smaller than predicted by the hydrostatic equilibrium, if the system behaves like a granular medium or if the object is not homogeneous (differentiated). These are two scenarios to explain the low density of Haumea compared to the hydrostatic equilibrium one for a homogeneous body, as shown in \citet{Ortiz2017}. In other words, the same reasons that make Haumea less dense than expected could be at play for 2002~TC$_{302}$. Therefore, in summary, either 2002~TC$_{302}$ is governed by granular physics (and/or it is differentiated) or the rotation of the body is slower than 5.4 \si{\hour}. We think the latter possibility is more plausible.
On the other hand, as already mentioned, it is well known that the scientific literature is biased toward short periods \citep{Sheppard2008} and in order to look for long rotation periods, extensive datasets are needed.

Hence, we combined all our runs in order to look for long periods in the data. To combine all our runs from 2014 to 2019, we did not use absolute photometry because it is often extremely difficult to achieve 0.02mag accuracy in the absolute calibrations using standard Johnson or Sloan filters, and we would still have to correct for solar phase angle effects. The absolute calibrations are even more problematic with unfiltered observations, which was our case (in order to achieve high signal to noise ratio). For all the above, the use of absolute calibrations results in jumps of several cents of magnitude from run to run, which is unacceptable to derive low amplitude light curves. Hence, we normalized the fluxes of each campaign to the mean value of the run. In other words, the fluxes were divided by the mean flux of the run. For long runs this method should work properly because the mean flux in the run should be similar to the mean flux averaged over a rotation cycle, but for short runs or for very long rotation periods, this method can introduce small shifts in the photometry and spurious frequencies in the periodograms. In our case, all runs lasted more than 3 days so the way of combining the runs by normalizing to the mean value is much better than using absolute calibrations, but we cannot completely discard that some effects are present.

\begin{figure}
\centering
	\includegraphics[width=\columnwidth]{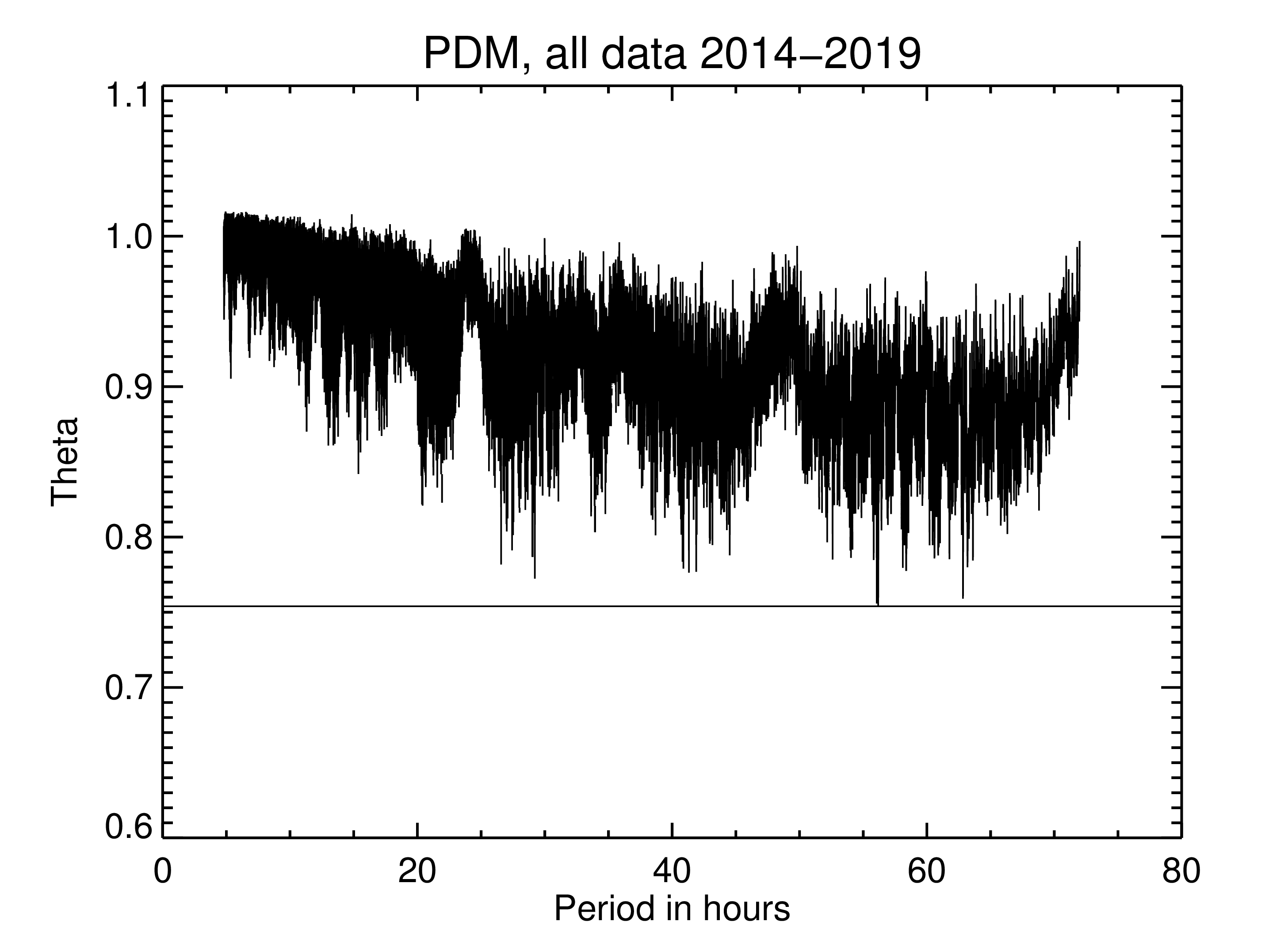}
    
    \caption{Phase Dispersion Minimization results for the entire photometry data set from 2014 to 2019. The horizontal line indicates where the minimum value is obtained, which corresponds a period of $\sim$56.1 h}
    \label{PDMphotometry}
\end{figure}

\begin{figure}
\centering
\includegraphics[width=\columnwidth]{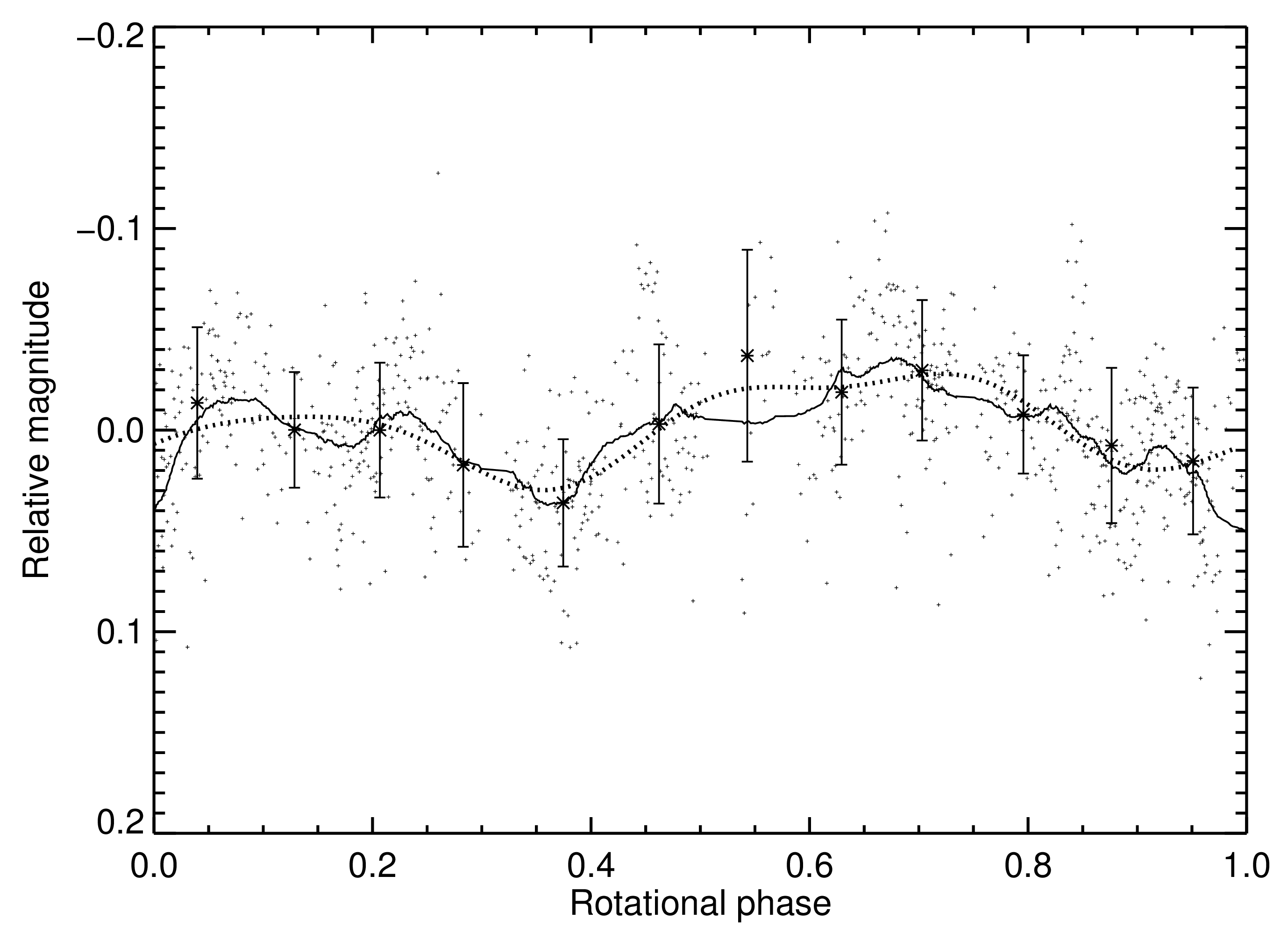}
    
    \caption{Photometry measurements of 2002 TC$_{302}$ folded to a period of 56.0642h. The small dots correspond to the single data points whereas the asterisks represent the median values binned in phase bins of 0.083. The error bars show the dispersion of the data in each bin. The solid line represents a smoothed curve using a width of 80 points and the dashed line shows a fourth order Fourier fit to the data. The light-travel corrected epoch for zero phase is JD 2456895.252777}
    \label{LightCurve}
\end{figure}

Once we combined all the datasets we analyzed all the photometry (with times corrected for light travel time) from 2014 to 2019 using the Phase Dispersion Minimization (PDM) technique. The PDM technique has the advantage over other period-finding techniques that a period is found independently of the shape of the light curve, and it is therefore more robust to finding double-peaked light curves caused by shape effects than other techniques. We found that a period of $\sim$56.1 h gives a clear minimum (see Figure \ref{PDMphotometry}). This period corresponds to a shape-induced rotational light curve because a relative minimum at $\sim 28$ h (half the best period) is also seen in the PDM plot.  In the plot, there is also a sharp minimum at $\sim64$ h, but it is considerably less pronounced so our preferred period is $\sim$56.1 h. From the analysis with the PDM technique it appears that we have a shape-induced rotational light curve of low amplitude with a period of $\sim$56.1 h. The peak to valley amplitude of a four order Fourier fit to the light curve folded to 56.1 h is $0.06\pm0.01$ mag (see Figure \ref{LightCurve}). This period is consistent with the period found in the astrometry residuals (explained in the next section) and therefore it would be consistent with the orbital period of a potential satellite whose spin and orbit would be locked. But it is possible that we have two or more photometric periods superimposed in the light curve (from the satellite and from the primary), making the analysis of this low-amplitude light curve even more complicated and making the identification of just one period difficult. 

\section{Time series astrometry}
\label{Astrometry}

The same large image dataset that was obtained for photometry purposes was also analyzed astrometrically with the same tools and techniques described in section \ref{Occultation predictions}. The astrometry was derived with respect to the Gaia DR2 catalog because it was already available when we started this analysis (but not at the time when the occultation preditions were made).  The complete astrometry dataset from 2014 to 2019 is presented as online material. An orbital fit to all the data was carried out and the residuals to the orbit were obtained both in Right Ascension and Declination.

The standard deviation of the residuals is 0.06 arcsec. This is somewhat larger than what can be expected given the signal to noise ratio of the observations. So this can mean that there are systematic errors or there are real short-term or long-term oscillations in the data. If the oscillations are real, they can be tied to the presence of a satellite. 

\begin{figure}[t]
\centering
\includegraphics[width=\columnwidth]{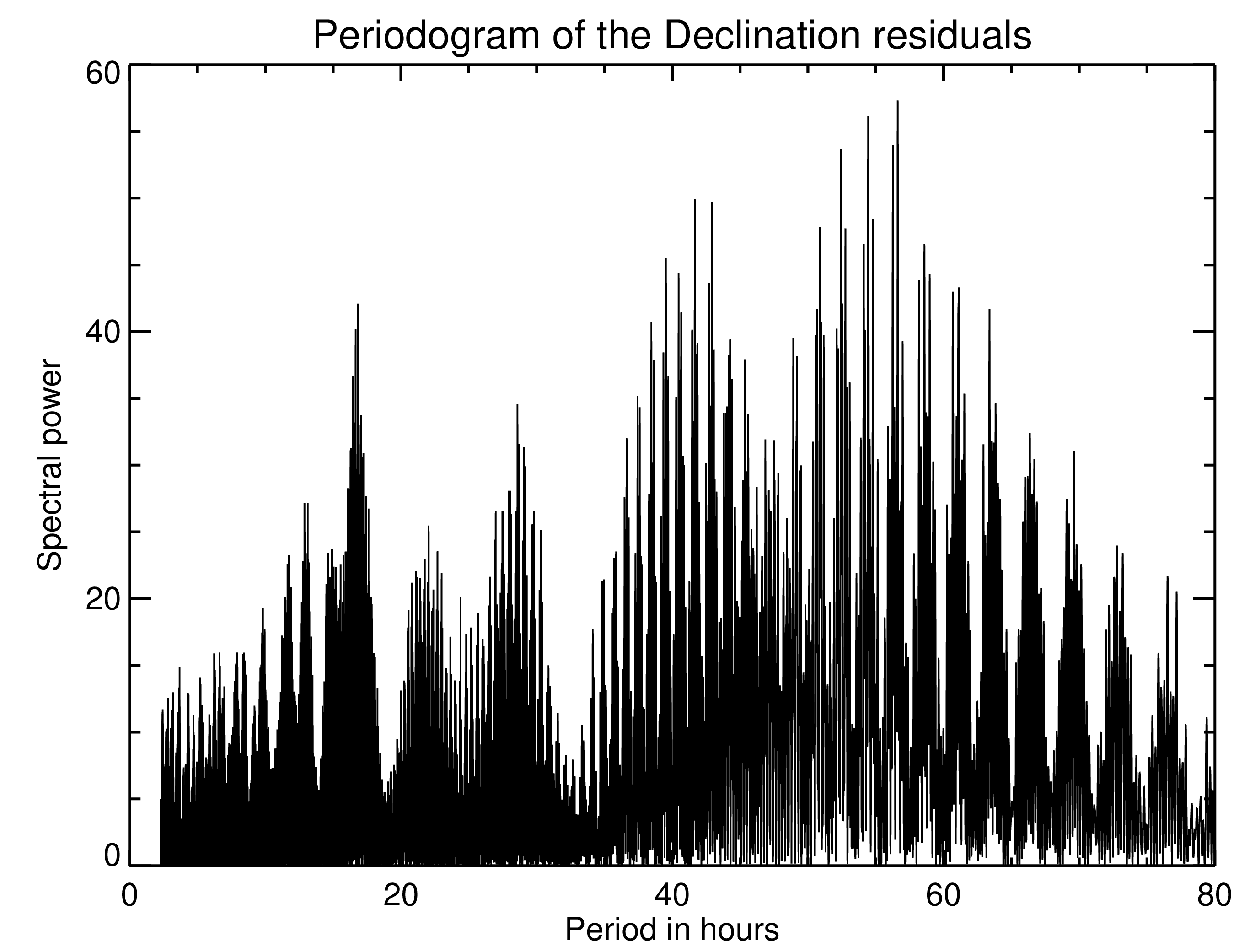}
    \caption{Lomb-Scargle periodogram of the  Declination residuals of an orbital fit to the astrometry measurements of 2002 TC$_{302}$. }
    \label{Lomb-residuals}
\end{figure}

\begin{figure}[t]
\centering
\includegraphics[width=\columnwidth]{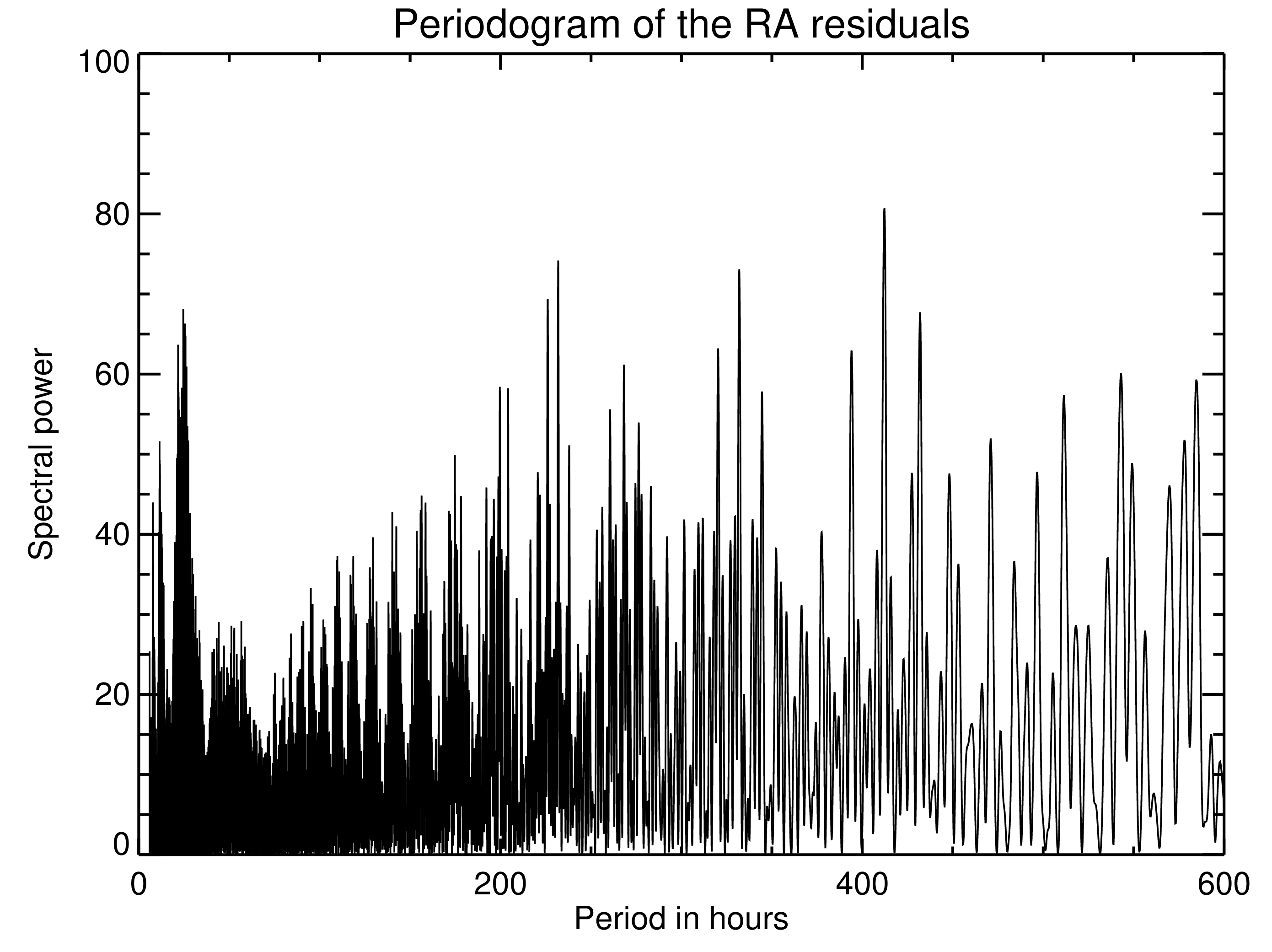}
    \caption{Lomb-Scargle periodogram of the Right Ascension residuals of an orbital fit to the astrometry measurements of 2002 TC$_{302}$. }
    \label{Lomb-residuals-RA}
\end{figure}

The epochs of the residuals were corrected for light travel time and the residuals were analyzed using the Lomb-Scargle periodogram technique as done for Orcus to reveal the oscillation caused by its satellite Vanth \citep{Ortiz2011}. Depicted in Figure \ref{Lomb-residuals}, the periodogram of the declination residuals shows three main peaks at 0.4239, 0.4265, and 0.4408 cycles/day (56.61, 56.27, and 54.44 h, respectively). The spectral power of these peaks are well above any other and their significance level is well above 99.9\%. The peak at 0.4239 cycles/day has a higher spectral power than the other two peaks but the other two cannot be ruled out, so it is clear that there is a periodic signal with frequency in the range 0.4239 to 0.4408 cycles/day, or periods between 54.4 and 56.6 h. In right ascension (RA) the periodogram of the residuals (Figure \ref{Lomb-residuals-RA}) shows peaks at 0.4233 and 0.4260 cycles/day although the highest peak in the periodogram corresponds to a frequency of 0.05823 cycles/day or a period of 412.1 h (17.17 days). If such a 17.17-day oscillation were caused by a satellite, the satellite would have been easily spotted in Hubble Space Telescope (HST) images (see discussion), so we tend to believe that it is an artifact because the residuals in RA showed larger scatter and might be more affected by differential chromatic refraction than those in declination. 


It therefore appears that the residuals have a periodicity in the range 0.4239 to 0.4408 cycles/day. Although the exact frequency/period is difficult to determine within this interval, our preferred one is 0.4239 cycles/day (56.61 h), which is very similar the period found with the Phase Dispersion Minimization (PDM) technique in the photometric datasets (see section \ref{Size and shape}). The peak to valley amplitude of a sinusoidal fit to the RA residuals folded to the 56.61 \si{\hour} period is 0.017 $\pm$ 0.006 arcsec. The peak to valley amplitude of a sinusoidal fit to the declination residuals phased to the 56.61 \si{\hour} period is 0.009 $\pm$ 0.003 arcsec. 

All this phenomenology is consistent with the idea that a satellite could be causing oscillations in the position of the photocenter. This is analyzed in some detail in the discussion (section \ref{Discussion}). 
In principle, the study of the phase of the RA and the phase of the Declination errors (when folded to the orbital period) could potentially give an idea of the orbit orientation of the satellite. If the orbit is more face-on, the RA and declination residuals should be out of phase with one another. For instance, a circular, clockwise, face-on orbit would first show the maximum of the Declination residuals, followed by the minimum in the RA residuals and after that the minimum of Declination would follow. Finally, the maximum in Right Ascension residuals would be reached. If the orbit is more edge-on, the RA and Declination residuals would be more in phase. Unfortunately, in our case we do not know the exact orbital period and the phases change dramatically depending on the period.
Note that the 56.61 \si{\hour} period is close, but not an exact match to the $\sim$ 56.1 \si{\hour} period in the photometry. We would expect the two periods to be identical if the putative satellite has its spin locked to its orbit.


\section{Discussion}
\label{Discussion}


Not counting the Pluto-Charon system, this is the best observed occultation by a TNO in terms of the number of chords, published in the literature thus far. The occultation by the dwarf planet Haumea had set a record the year before to this occultation \citep{Ortiz2017}. More recently, a stellar occultation by the TNO Huya was observed and the number of chords was even larger than for the case of 2002~TC$_{302}$, but the analysis of the Huya results is still ongoing and only preliminary results have been presented in \cite{Santos-Sanz2019}. The large number of chords obtained on 2002~TC$_{302}$ has allowed us to determine size and shape accurately.

From Herschel and Spitzer thermal measurements, \cite{Fornasier2013} derived an equivalent-area diameter of $584.1^{+105}_{-88}$ \si{\kilo\meter} for 2002~TC$_{302}$  with uncertainties at the 1 $\sigma$ confidence level. This is around 84 \si{\kilo\meter} larger than the value derived here from the occultation $(543\times460)^{1/2}=499.8$ \si{\kilo\meter}. Even though the values are compatible within the large 88 \si{\kilo\meter} error bar of the thermal measurements  (which is only at the 1 $\sigma$ level), it is pertinent to note that thermal models tend to underestimate the real effective diameters for ellipsoidal bodies, specially for those with high obliquity, as pointed out by \cite{Brown1985} and as clearly demonstrated for Haumea \citep{Ortiz2017}. In other words, we were expecting that the occultation would give a larger equivalent-area diameter than that of the thermal models, but found the opposite.
Therefore, the difference of $\sim84$ \si{\kilo\meter} could be even more significant. 
The difference can potentially be explained by the existence of a satellite. Note that the equivalent-area diameter of a binary would be $(543 \times 460) + D_s^2 = 584^2$, where $D_s^2$ is the equivalent-area diameter of the putative satellite. When solving for $D_s^2$ we come up with $D_s=302$ \si{\kilo\meter}.  An upper limit to the diameter for the putative satellite can be obtained from $(543\times460) + (D_s^{max})^2 = (584+105)^2$. The resulting maximum diameter for the satellite $D_s^{max}$ would be 474 km, which is almost as large as the main body and this is probably too large to have been undetected in the occultation. If the albedo of the potential satellite was smaller than that of the primary, a somewhat smaller satellite diameter would be possible while still giving the thermal output modeled in \cite{Fornasier2013}. For that reason we are giving a plausible size range of 100 \si{\kilo\meter} to 300 \si{\kilo\meter} for the potential satellite.
 It is also worth mentioning that the 88 km error bar of the thermal diameter reported in \cite{Fornasier2013} is strikingly large compared to other error bars determined for TNOs of similar size. When looking at the fit to the spectral energy distribution of 2002~TC$_{302}$ in that paper (their figure 15), it turns out that the fit is poor. 
According to Kiss (2019, priv. comm.) the Herschel PACS fluxes used in \cite{Fornasier2013} came from the combination of observations at two epochs, but a close look at the Herschel PACS images of the second epoch revealed contamination from a bright source. If only the uncontaminated epoch is used to derive the fluxes, the fit improves considerably, decreasing the error bars and the fitted effective diameter is somewhat larger. With this information the difference from the thermal diameter and the occultation diameter is probably significant at more than 2$\sigma$.

The existence of a satellite of at least around 100 \si{\kilo\meter} in size and up to $\sim$300 \si{\kilo\meter}, close to the body, might explain at least part of the difference in size with the thermal measurements and could also explain the oscillations in the high-accuracy time series astrometry as well as the photometry variability. 

 Regarding the photometry variability, if 2002~TC$_{302}$ has a satellite of $>$ 100 \si{\kilo\meter} in size, with a similar albedo to the main body, the contribution of the satellite to the total brightness of the system would be $>$(100/499.8)$^2$ in percentage, or around $\gtrsim$0.05 mag. A satellite with a diameter of 200 \si{\kilo\meter} would be capable of contributing 0.16 mag, and a satellite of 300 \si{\kilo\meter} would give rise to a 0.26 mag contribution, so rotational modulation of the satellite would have contributions below 0.05, 0.16 and 0.26 mag respectively and might explain the low amplitude of the light curve. Hence, we believe that an unresolved satellite could account for a large part of the short-term variability observed. Given that a satellite with 30\% of variability due to shape irregularities and 200 km in size would produce oscillations of $0.3\times0.16\%$ or around 0.05 mag, which is not far from the 0.06 mag variability observed, the scenario of a $\sim200$ km satellite producing the observed light curve looks coherent. Note that TNOs in the 200 km size range may already be too small to have hydrostatic-equilibrium dominated shapes and there is some evidence that small TNOs have higher light curve amplitudes than the larger TNOs. In fact there is a correlation of light curve amplitude with absolute magnitude \citep{Sheppard2008}.


A potential scenario to explain our photometric observations would be that the putative satellite gives rise to the $\sim$56.1 \si{\hour} period whereas the main body would have been slowed down by the tidal interaction with the satellite and be an oblate spheroid with low variability, which would be difficult to detect and disentangle from the longer period. The periodicity of $\sim$56.1 h seems consistent with the orbital period of an unresolved satellite from the astrometry residuals, although the two periods do not exactly match. Note that the light curve would be induced by the shape of the satellite, not by eclipses.

According to the different expressions for the tidal locking timescales applied to TNOs with satellites in \cite{Thirouin2014} and \cite{Fernandez-Valenzuela2019}, the putative satellite would have synchronized its orbit and spin in times of the order of one to one hundred million years, assuming typical values for the tidal dissipation parameter Q and for the rigidity of ice, and assuming a satellite radius of 100km and a density of 600 kg/m$^3$. This timescale is orders of magnitude lower than the age of the solar system. Hence, it appears likely that the putative satellite would be tidally locked, although a caveat must be made in the sense that most of the tidal timescale expressions often rely on oversimplifications of the complex physics of the tidal interaction \citep{EfroimskyWilliams2009}. Therefore we cannot discard that the rotation period of the potential satellite could be different than the orbital period. 

Regarding the rotation period of the primary, as mentioned before, we could not firmly determine it. It appears possible that the primary rotation could also be tidally locked with the satellite orbital period, but in that case, a large body such as 2002~TC$_{302}$, presumably in hydrostatic equilibrium or close to it, spinning at $\sim$56.1 h would have an oblate shape with axial ratios close to 1. This is far from the observed 1.18 value for the projected axial ratio seen in the occultation. Hence, a faster spin period than $\sim$56.1 h for the primary seems to be required. If the primary rotated at half that period the axial ratio of a Maclaurin body with a density around 800 kg/m$^3$ would be 1.02, still considerably below 1.18, so this possibility also seems incompatible with the occultation observations and a shorter rotation period for the primary would be favored.

The putative satellite would have to be very close to the main body so that HST observations did not reveal it. There are only 2 images of 2002~TC$_{302}$ in the HST archive from which no satellite has been reported. We know that HST cannot resolve binary objects that are separated less than at least several tens of mas, which is the diffraction limit of the telescope. Therefore, any satellite orbiting at less than $\sim 2000$ \si{\kilo\meter} from the main body, would be challenging to detect.

Assuming that the density of 2002~TC$_{302}$ is around 800 \si{\kilo\gram\per\meter\cubed}, which is the density expected for a TNO of this size \cite[according to the plots shown in, e.g.,][]{Grundy2015,BiersonandNimmo2019,Grundy2019}, the mass of the central body can be estimated. This allows us to compute the distance at which a putative satellite would have an orbital period of $\sim$56.1 \si{\hour}. This distance would be $\sim$ 1780 \si{\kilo\meter}. 
Given that 2002~TC$_{302}$ is currently at 43 au, a semiaxis of 1780 \si{\kilo\meter} implies 0.058 arcsec. This is the maximum angular separation and would already be challenging to resolve with HST. During most of the parts of the orbit the angular separation would be well below the resolution limit for HST, but it may be possible to resolve using the Near Infrared Camera on board the James Webb Space Telescope, which has a pixel scale of 0.031 arcsec/pixel at short wavelengths (0.6 to 2.3 \si{\micro\meter}).

If the putative satellite has an effective equivalent-area diameter of 200 \si{\kilo\meter} and that of the main body is 499.8 \si{\kilo\meter}, the ratio of areas and therefore the ratio of brightness is (200/499.8)$^2$ = 0.16. Hence, the photocenter would lie at a distance of $0.058 \times 0.16 = 0.009$ arcsec from the central body and this means that the total oscillation would be twice that value, or 0.018 arcsec. This is close to the fitted amplitude of the oscillation of the astrometric residuals. A small correction must be applied because the photocenter rotates around the barycenter and the main body is not exactly at the barycenter, but 0.004 arcsec away from it. Therefore, the expected amplitude of the residuals would be 0.014 arcsec, which is even closer to the observational results. This derivation assumed equal albedo for the satellite and the main body. The reality may be somewhat different, so the needed diameter for the satellite may not be exactly 200 \si{\kilo\meter}, depending on the exact value of its geometric albedo.

A close satellite would have also been capable of slowing down the rotation of 2002~TC$_{302}$ through tidal interaction. This would also explain that 2002~TC$_{302}$ could have adopted a nearly oblate shape instead of a more triaxial shape as is the case for 2003~VS$_2$, which is a very similar body to 2002~TC$_{302}$ in terms of size \citep[its triaxial shape has been determined using the same techniques presented in this work;][]{Benedetti2019}. A slow rotation of $\sim20$ \si{\hour} could be explained by the tidal interaction of, for instance, a $\sim 200$ \si{\kilo\meter} satellite orbiting at $\sim$ 1800 \si{\kilo\meter} from the primary. 

In this regard, it is pertinent to analyze the system in terms of the specific angular momentum (angular momentum divided by $(G M^3 R)^{1/2}$ where $G$ is the gravitational constant, $M$ the mass and $R$ the radius of the body).
The specific total angular momentum ($H$) of a system formed by a primary and a satellite 
(understood as the sum of the orbital angular momentum plus the angular momentum of the primary and that of the satellite)
was computed according to the following equations from \cite{DescampsMarchis2008}:
\begin{multline}
H =\frac{q}{(1+q)^{\frac{13}{6}}}\sqrt{\frac{a(1-e^{2})}{R_{p}}} + \frac{2}{5} \frac{\lambda_{p}}{(1+q)^{\frac{5}{3}}} \Omega + \\
\ \frac{2}{5} \lambda_{s} \frac{q^{\frac{5}{3}}}{(1+q)^{\frac{7}{6}}}\left (\frac{R_{p}}{a} \right)^{\frac{3}{2}}
\end{multline}
where {\it q} is the secondary-to-primary mass ratio, {\it a} the semimajor axis, {\it e} the eccentricity, and {\it R$_{p}$} the primary radius.
The {\it $\Omega$} parameter is the normalized spin rate expressed as:
\begin{equation}
\Omega = \frac{\omega_{p}}{\omega_{c} }
\end{equation}
where $\omega_{p}$ is the primary rotation rate and $\omega_{c}$ the critical spin rate for a spherical body:
\begin{equation}
\omega_{c} = \sqrt{\frac{GM_{p}}{{R_{p}^3}}};
\end{equation}
here {\it G} is the gravitational constant and {\it M$_{p}$} the mass of the primary.
Assuming a triaxial primary with semi-axes as $ a_{o}> a_{1} > a_{2}$, the $\lambda_{p}$ shape parameter is
\begin{equation}
\lambda_{p}= \frac{1+\beta^{2}}{2(\alpha\beta)^\frac{2}{3}}
\end{equation}
where $\alpha$ = $a_{2}/a_{0}$ and $\beta$ = $a_{1}/a_{0}$. \\
In this work, we considered the primary to be nearly spherical ($\lambda_{s}$=1.0) and the satellite to be somewhat nonspherical with $\lambda_{s}$=1.2\\
\newline

 Using those expressions, the specific angular momentum of
a body of the size of 2002~TC$_{302}$ (with an expected density around 800 \si{\kilo\gram\per\meter\cubed}) spinning at a primordial $\sim$ 7.7 \si{\hour} period would  be $\sim$ 0.2. The primordial rotation period is taken from the Maxwellian fit to the rotation periods of TNOs presented in \cite{Duffard2009}.
On the other hand, a body with a rotation period of 20 \si{\hour} would have a much smaller specific angular momentum than 0.21, but if 2002~TC$_{302}$ rotates at 20 \si{\hour} and has, for instance, a tidally locked satellite with a mass ratio of 0.065 to the primary and orbiting at 1780 \si{\kilo\meter} from the main body, the specific angular momentum would be 0.21, meaning that such a satellite could have slowed down 2002~TC$_{302}$ from a primordial spin to a rotation period of $\sim20$ \si{\hour}, conserving the total angular momentum of the system. Note that a body with a size around 200 \si{\kilo\meter} and with the same density as the central body would have a mass ratio of 0.065  (which is the $q$ parameter that enters the specific angular momentum expression). If the density of the satellite is somewhat smaller than that of the primary, a slightly larger size would be needed for the satellite to give the required angular momentum. Other satellite configurations with smaller mass ratios are possible, and would have slowed down the primary to a period smaller than 20h. For instance, a satellite with a 145 \si{\kilo\meter} diameter would give a mass ratio around 0.024 and would have slowed down the primary to $\sim$ 10 \si{\hour} if orbiting at 1780 \si{\kilo\meter}. As a summary, a satellite with a size ranging from $\sim$ 145 to $\sim$ 200 \si{\kilo\meter} orbiting at $\sim$ 1800 \si{\kilo\meter} seems to offer a good overall fit to the phenomenology observed. 

Using the projected area of the occultation ellipse and the absolute magnitude of 2002~TC$_{302}$ \citep[H$_V$=4.23;][]{Tegler2016}), we can derive the geometric albedo as done in, e.g., \cite{Ortiz2017} for Haumea. The resulting value is 0.147 for 2002~TC$_{302}$. Comparing the geometric albedo of 2002~TC$_{302}$ to those of similar size TNOs for which stellar occultations have been observed (so that the size is accurately known and thus the geometric albedo can also be derived with high accuracy), it turns out that 2002~TC$_{302}$ has a higher geometric albedo than that of 2003~AZ$_{84}$ \citep[after accounting for its known satellite as shown in][]{Ortiz2020}, a slightly higher albedo than that of 2003~VS$_2$ \citep{Benedetti2019} and also slightly higher than that of G!k\'un$||$'h\`omd\'im\`a (2007~UK$_{126}$) \citep[also accounting for its known satellite,][]{Ortiz2020}. Also, the geometric albedo of 2002~TC$_{302}$ is higher than that of Quaoar, which is considerably larger than 2002~TC$_{302}$. The somewhat higher albedo than expected might also be a hint that 2002~TC$_{302}$ could have a large satellite because in that case, we would be using the H$_V$ corresponding to the combination of the satellite and the main body, whereas we should use a larger H$_V$ value, which would decrease the geometric albedo computation. A satellite of 200 \si{\kilo\meter} with a similar albedo to the primary would contribute around 16\% of the brightness, so the geometric albedo would be around 16\% smaller or around 0.127. This value is closer to the geometric albedo determined by Herschel and Spitzer measurements \citep{Fornasier2013} and closer to that of similar size TNOs. Nevertheless, we cannot expect that the geometric albedo of the TNOs depends only on size. Different formation processes in different areas of the solar nebula may imply different surface compositions, and also posterior dynamical and evolution processes including collisions may have important effects on the final albedo of a TNO. Therefore, the argument on the albedo is just another hint but cannot be considered as conclusive evidence. According to \cite{Barkume2008}, 2002~TC$_{302}$ has water ice spectroscopically detected and since water ice is highly reflective, it makes sense that 2002~TC$_{302}$ could be somewhat brighter than the average TNO (for non-water-ice-bearing bodies), but note that 2003~VS$_2$, 2007~UK$_{126}$, 2003~AZ$_{84}$ and Quaoar also have indications of water ice in their spectra \citep{Barucci2011} whereas their geometric albedos are lower than for 2002~TC$_{302}$.
In the context of the Uranian satellites it is well known that they show strong water ice bands, and have a variety of moderately low albedos, starting just a little higher than that of 2002~TC$_{302}$ \citep[e.g.][]{Buratti1991} that depend on the mixture of water ice with pollutants. There is also abundant laboratory and modeling literature that describes how the visible albedo and near-infrared absorption band depths in granular water ice are affected by admixture of dark pollutants \citep[e.g.][]{Clark1984}.

While there are clear indications for a satellite, none of the 12 occultation light curves have detected the putative satellite. Given that the inter-spacing of the chords is smaller than 200 \si{\kilo\meter}, the putative satellite should have been detected if it were sufficiently close to the main body at the time of the occultation. But the sampling of the chords is not good enough away from the main body in the cross track direction, so a satellite could have been easily missed. We know that Huya, a TNO of similar size to 2002~TC$_{302}$, has a large and close satellite of  213$\pm$30 km in diameter, as estimated from unresolved thermal observations \citep{Fornasier2013}, but this satellite has not been detected in the stellar occultation preliminary reported in \cite{Santos-Sanz2019}, despite a large number of chords. This clearly illustrates that the putative satellite of 2002~TC$_{302}$ could have easily gone undetected.

Apart from no satellite detection, no ring features or dust structures have been detected during the occultation either. From the occultation dataset with the least scatter, an upper limit to the width of a Chariklo-like dense ring is 7 \si{\kilo\meter} at 3$\sigma$. This means that a ring similar to that observed at Chariklo would have been missed, but a ring similar in width to that of Haumea would have been detectable. An intermediate ring between that of Chariklo and Haumea, in terms of width, would also have been detected if it existed.





\section{Conclusions}
\label{conclusion}

2002~TC$_{302}$ caused a stellar occultation on 28$^{th}$ January 2018 from which its projected size and shape at the time of the occultation could be derived with high accuracy. Not counting the Pluto-Charon system, this is the best observed occultation by a TNO in terms of the number of chords published in the scientific literature thus far. The elliptical fit to the 12 chords has a major axis of $543\pm18$ \si{\kilo\meter} and minor axis of $460\pm11$ \si{\kilo\meter}. This implies an equivalent-area diameter of 499.8 \si{\kilo\meter} and a geometric albedo of 0.147 (for an absolute magnitude of 4.32). The smaller equivalent diameter than the radiometrically derived value from Herschel and Spitzer observations and the larger albedo could be hints for the presence of a large satellite close to 2002~TC$_{302}$. There are other hints from ground based observations that point in that direction. 
From the sharp disappearance and reappearance of the star in the occultation, we can conclude that 2002~TC$_{302}$ lacks a global atmosphere, with an upper limit of the order of 100 nbar. No ring features or dust structures were detected close to the nucleus, although the data lacked the needed quality to discover a dense ring of the width of Chariklo's. However, a dense ring of the width of that of Haumea would have been easily detected. An intermediate ring in terms of width would also have been detected if present.

\begin{acknowledgements}
 This research was partially based on data taken at the Sierra Nevada Observatory, which is operated by the Instituto de Astrof\'isica de
Andaluc\'ia (CSIC). This research is also partially based on data taken at the German-Spanish Calar Alto observatory, which is jointly
operated by the Max Planck Institute für Astronomie and the Instituto de Astrof\'isica de Andaluc\'ia (CSIC). Part of the results were also based on observations taken at the 1.6m telescope on Pico dos Dias Observatory. This research was partially based on observations collected at the Schmidt telescope 67/92 cm  (Asiago, Italy) of the INAF - Osservatorio Astronomico di Padova.
Funding from Spanish projects
AYA2014-56637-C2-1-P, AYA2017-89637-R, from FEDER, and Proyecto de Excelencia de la Junta de Andaluc\'ia 2012-FQM1776 is acknowledged. 
We would like to acknowledge financial support by the Spanish grant AYA-RTI2018-098657-JI00 ``LEO-SBNAF'' (MCIU/AEI/FEDER, UE) and the financial support from the State Agency for Research of the Spanish MCIU through the ``Center of Excellence Severo Ochoa'' award for the Instituto de Astrof\'isica de Andaluc\'ia (SEV- 2017-0709). Part of the research received funding from the European Union's Horizon 2020 Research and Innovation Programme, under grant agreement no. 687378 and from the ERC programme under Grant Agreement no. 669416 Lucky Star.
The following authors acknowledge the respective CNPq grants: FB-R 309578/2017-5; RV-M 304544/2017-5, 401903/2016-8; J.I.B.C.
308150/2016-3; MA 427700/2018-3, 310683/2017-3, 473002/2013-2. This study was financed in part by the Coordenação de Aperfeiçoamento de Pessoal de Nível Superior - Brasil (CAPES) - Finance Code 001 and the National Institute of Science and Technology of the e-Universe project (INCT do e-Universo, CNPq grant 465376/2014-2). GBR acknowledges CAPES-FAPERJ/PAPDRJ grant E26/203.173/2016, MA FAPERJ grant
E-26/111.488/2013 and ARGJr FAPESP grant 2018/11239-8.
E. F-V. acknowledges support from the 2017 Preeminent Postdoctoral Program (P$^3$) at UCF.
C.K., R.S., A.F-T., and G.M. have been supported by the K-125015 and GINOP-2.3.2-15-2016-00003 grants of the Hungarian National Research, Development and Innovation Office (NKFIH), Hungary.
G.M. was also supported by the Hungarian National Research, Development and Innovation Office (NKFIH) grant PD-128360. R.K. and T.P. were supported by the VEGA 2/0031/18 grant.
We acknowledge the use of Occult software by D. Herald.
\end{acknowledgements}

%
%

\bibliographystyle{aa}
\bibliography{bibtex_2002TC302}

\end{document}